\documentclass[12pt]{article}

\usepackage{amsmath}
\usepackage{graphicx}
\usepackage{cite}

\newcommand{\mys}[1]{\section{#1}
   \hspace{0.8cm}\setcounter{equation}{0}}
\renewcommand{\theequation}{\arabic{section}.\arabic{equation}}
\newcommand{\myappendix}{\appendix
   \renewcommand{\theequation}{\Alph{section}.\arabic{equation}}
   \vspace{30pt} \noindent {\Large \bf Appendix}}

\newlength{\dummysp}
\newcommand{\diag}{\mathop{{\hbox{diag} \, }}\nolimits}
\newcommand{\tr}{\mathop{{\hbox{Tr} \, }}\nolimits}

\newcommand{\half}{\frac{1}{2}}
\newcommand{\third}{\frac{1}{3}}

\newcommand{\beq}{\begin{eqnarray}}
\newcommand{\eeq}{\end{eqnarray}}
\newcommand{\nnn}{ \nonumber \\ }
\newcommand{\p}{{\partial}}
\newcommand{\e}{{\epsilon}}
\newcommand{\s}{{\sigma}}

\newcommand{\ord}[1]{{{\cal O}(#1)}}
\newcommand{\gappeq}{\mathrel{\rlap {\raise.5ex\hbox{$>$}}
{\lower.5ex\hbox{$\sim$}}}}
\newcommand{\lappeq}{\mathrel{\rlap{\raise.5ex\hbox{$<$}}
{\lower.5ex\hbox{$\sim$}}}}
\newcommand{\myref}[1]{(\ref{#1})}

\newcommand{\ben}{\begin{enumerate}}
\newcommand{\een}{\end{enumerate}}
\newcommand{\bit}{\begin{itemize}}
\newcommand{\eit}{\end{itemize}}
\newcommand{\Cbf}{{\bf C}}
\newcommand{\Rbf}{{\bf R}}

\newcommand{\fourth}{\frac{1}{4}}

\newcommand{\ghat}{{\hat g}}

\newcommand{\Ncal}{{\cal N}}
\newcommand{\adsv}{{AdS}_5}
\newcommand{\tp}{{\theta_+}}
\newcommand{\tm}{{\theta_-}}

\newcommand{\rhat}{{\hat r}}
\newcommand{\Rhat}{{\hat R}}
\newcommand{\phii}{{\phi_i}}
\newcommand{\phij}{{\phi_j}}

\def\[{\left [}
\def\]{\right ]}
\def\({\left (}
\def\){\right )}

\begin{document}

\begin{titlepage}

\renewcommand{\thefootnote}{\fnsymbol{footnote}}

\pagestyle{empty}
\begin{flushright}
UMN-TH-2502/06 \\
FTPI-MINN-06/15 \\
CERN-PH-TH/2006-095\\
hep-th/0605212 \\
August 2006
\end{flushright}
\vspace*{5mm}

\begin{center}
{\bf \Large Bulk fields in $\adsv$ from probe D7 branes}
\end{center}

\vspace{0.02cm}
\begin{center}
{\sc Tony Gherghetta\footnote{{\tt tgher@physics.umn.edu}}$^{a,b}$} 
{\small and}
{\sc Joel Giedt\footnote{{\tt giedt@physics.umn.edu}}$^{c}$}
\end{center}

\begin{center}

$^a${\it\small School of Physics and Astronomy, 
University of Minnesota,\\ Minneapolis, MN 55455, USA}

$^b${\it\small Theory Division, CERN, CH-1211 Geneva 23, Switzerland}

$^c${\it\small  William I. Fine Theoretical Physics Institute,
University of Minnesota, \\Minneapolis, MN 55455, USA}

\end{center}

\vspace{0.5cm}
\begin{abstract}
We relate bulk fields in Randall-Sundrum $\adsv$ phenomenological
models to the world-volume fields of probe D7 branes in the Klebanov-Witten 
background of type IIB string theory.  The string constructions are described 
by $\adsv\times T^{1,1}$ in their near-horizon geometry, with 
$T^{1,1}$ a 5d compact internal manifold that yields $\Ncal=1$ supersymmetry
in the dual 4d gauge theory. The effective 5d Lagrangian description derived
from the explicit string construction leads to additional features that are not 
usually encountered in phenomenological model building.
\end{abstract}

\end{titlepage}

\renewcommand{\thefootnote}{\arabic{footnote}}
\setcounter{footnote}{0}


\vspace{0.5in}

\mys{Introduction}
Phenomenological models in a slice of $\adsv$ have recently provided
a new alternative to the usual four-dimensional (4d) supersymmetric models 
in addressing the hierarchy problem. These models
generalize the Randall-Sundrum model (RS1)~\cite{Randall:1999ee} 
by allowing for the presence of bulk fermion and gauge fields. 
A striking feature of these models is that they are conjectured to be dual to 4d 
gauge theories with a composite Higgs and top quark~\cite{Gherghetta:2006ha}. 
In all these bottom-up constructions the bulk fields are put in by hand. However 
one would like to seek a more fundamental description of these fields from a string 
construction where, in particular, the ``preons'' of the composite states are 
identified. In this article we begin a study to determine the possible
top-down constructions.

From the top-down there are various well-defined 
string constructions that, in the near-horizon geometry, 
are well-described by an $\adsv \times X_5$ background,
with $X_5$ a five-dimensional (5d) compact manifold.  
The simplest example is the one
that appeared in the seminal works on the AdS/CFT
correspondence, where $X_5=S^5$ \cite{Maldacena:1997re,
Gubser:1998bc,Witten:1998qj}. However, in the $X_5=S^5$
case the dual 4d gauge theory has $\Ncal=4$ supersymmetry,
which is too restrictive for phenomenological purposes.

\begin{figure}
\begin{center}
\includegraphics[height=3.5in,width=3.5in]{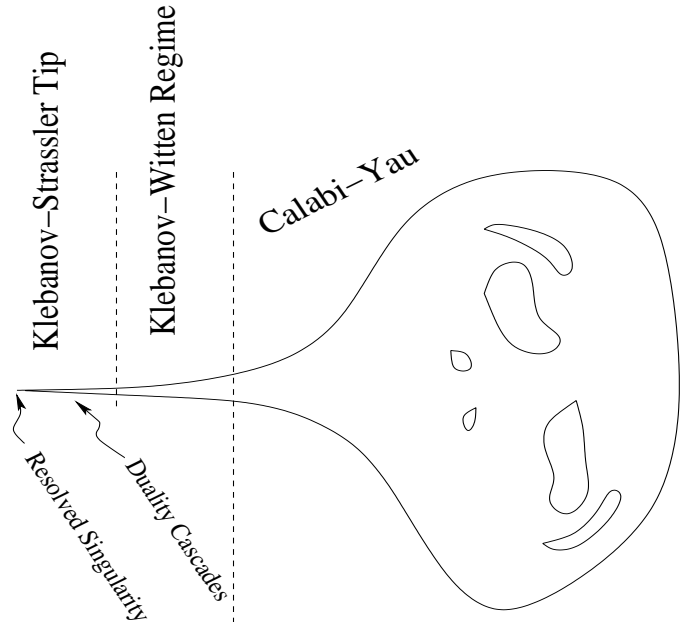}
\caption{A KS throat with a CY
glued on to regulate the UV of the dual gauge theory.
The KS tip serves to regulate the IR of the dual
gauge theory.  A section of
$\adsv$ is asymptotically contained in the throat.
The CY and the KS tip are, respectively, to be
thought of as refinements of the UV and IR branes
of RS1.  Much of the throat can be 
approximated by the significantly
simpler KW solution that we study in the text.
Furthermore, for the construction that we study,
the end of the D7 brane provides an alternative IR boundary,
as far as ``quark'' fields are concerned.
\label{mf1}}
\end{center}
\end{figure}

More interesting for our present purposes is the
Klebanov-Witten (KW) construction \cite{Klebanov:1998hh}, where
$X_5 = T^{1,1} \simeq (SU(2) \times SU(2)) / U(1)$.
In this case, the dual description is a 4d $\Ncal=1$ superconformal
gauge theory.  The theory is unregulated in the infrared (IR),
corresponding to a naked singularity in the geometric
picture.  The resolution of this singularity
leads to the more refined, Klebanov-Strassler (KS)
construction~\cite{Klebanov:2000hb}.  In both the
KW and KS backgrounds, the geometry is noncompact,
extending to infinity in the direction associated
with the $\adsv$ radius.  Correspondingly,
the ultraviolet (UV) of the dual gauge theory lacks a regulator.
On the gravity side of the duality,
this can be addressed if the geometry is completed
by a compact Calabi-Yau (CY) manifold in the region
far from the ``throat''.  When this is done in
the KS construction, as has been considered
by Giddings, Kachru and Polchinski~(GKP) \cite{Giddings:2001yu}, 
the spectrum is normalizable and discrete.
This is very much like the RS1 setup:
the tip of the KS throat represents the
IR brane, and the compact CY represents the
UV brane.  The general picture is sketched in Fig.~\ref{mf1}.

However, it is expected that the effective 5d action
describing the dimensional reduction of the string model
on $X_5$ will differ in a variety of ways from typical
phenomenological $\adsv$ actions.  Ultimately,
we want to compare the two classes of effective actions,
to delineate the extent to which they differ, and
to better understand the circumstances in which they agree.  
Furthermore, we would like to obtain string-inspired
constraints on the 5d phenomenological models.  
For instance, what 5d bulk masses are possible, what are the
boundary conditions at the branes, and what is the possible 
field content? In this article we address 
some of these questions.

The construction that is studied here involves
embedding a D7 probe brane in the KW geometry. In 
this way bulk matter and gauge fields are introduced; 
i.e., fields that are not already
contained in the $\Ncal=2$ $\adsv$ supergravity multiplet
associated with type IIB compactification on $T^{1,1}$
\cite{Gubser:1998vd,Jatkar:1999zk,Ceresole:1999zs,Ceresole:1999ht}.
In the dual 4d gauge theory, this corresponds to
the introduction of ``quark'' flavors.
The D7 brane wraps an internal cycle of the 
internal 5d compact space $T^{1,1}$ in such
a way that it ``disappears'' at some distance
from the tip of the throat, much as in
the $\adsv \times S^5$ models with probe branes
\cite{Karch:2000gx,Karch:2002sh,Kruczenski:2003be}.
This ending of the D7 brane leads to an IR boundary 
for the theory. In the dual 4d gauge theory, it has the effect
of bare masses for the quarks.  The IR cutoff
of the KS background corresponds to the confinement
scale of the dual gauge theory.  We will work
in the heavy quark limit, so that the flavors
are not propagating degrees of freedom near
the confinement scale.
Thus, for the investigation that follows, the
IR cutoff of KS is not a detail that we
will need.  For this reason, we will utilize the
simpler KW geometry.

Next we give a summary of the remainder of this article:
\bit
\item
In \S\ref{d7kw} we outline the bosonic fluctuations
associated with embedding a probe D7 brane in the
KW background, following Levi and Ouyang \cite{Levi:2005hh}.
We introduce a reparameterization of the embedding
coordinates that allows us to describe the action
of the embedding scalars as a function of the $\adsv$ radius.
We extract the effective 5d action that describes
the corresponding modes that are independent of the angular
coordinates of the internal 5d space $T^{1,1}$.
We briefly address the boundary terms that appear
for these scalars.  We find that a certain convenient
simplification occurs.  Finally, we summarize the
action of the D7 brane worldvolume vector boson,
and make some remarks on the effective 5d compactification.
\item
In \S\ref{r5d}, we isolate the regime in which the
scalar action can be mapped into conventional
$\adsv$ formulations, such as phenomenological 
models in a slice of $\adsv$.  We show that both bulk and
boundary masses arise.  The bulk mass-squared
is negative, but satisfies the Breitenlohner-Freed\-man
bound \cite{Breitenlohner:1982bm,Breitenlohner:1982jf}, 
just as in the cases of probe
branes in $\adsv \times S^5$ geometries 
\cite{Karch:2000gx,Karch:2002sh,Kruczenski:2003be}.
We describe how effects near the IR boundary
where the D7 brane ends can be
translated into an effective description
that resembles what is typically considered
in model-building applications.
In particular, we show how Cauchy boundary
conditions at the IR boundary of the effective
$\adsv$ regime are generic, and how they emerge
from Dirichlet boundary conditions that
occur at the radius where the D7 brane ends.
\item
In \S\ref{conc} we make concluding remarks and
outline some outstanding issues.
In particular, our present findings indicate
the need for a thorough understanding of
the harmonics on the somewhat complicated geometry
of the internal 3-manifold of the embedding,
and of the detailed effects near the end of
the D7 brane---items that we have touched on,
but mostly left to future work.
\item
In the Appendix, we discuss the geometry
and topology of the D7 embedding.  We show that
the 3-manifold wrapped by the D7 brane at fixed $\adsv$
radius is topologically equivalent to $S^3$.
This justifies an angle-independent assumption that
is made in the main text.  Of course this assumption
is only valid for fields that can be expanded on scalar harmonics 
of the 3-manifold.  We also identify the isometry 
group of the embedding.
\eit

\mys{D7 branes in the Klebanov-Witten background}
\label{d7kw}
Let us begin by reviewing the KW background~\cite{Klebanov:1998hh}. 
The ten-dimensional (10d) metric is given by
\beq
&& ds_{10}^2 = H(r)^{-1/2} \eta_{\mu \nu} dx^\mu dx^\nu
+ H(r)^{1/2} ds_{6}^2, \nnn
&& ds_{6}^2 = dr^2 + r^2 ds_{T^{1,1}}^2,
\quad H(r) = 1 + \frac{L^4}{r^4},
\eeq
where $L$ is the curvature length and $ds_{T^{1,1}}^2$ is the metric for $T^{1,1}$,
\beq
ds_{T^{1,1}}^2 = \frac{1}{9} \( d\psi + \sum_{i=1,2} \cos \theta_i
~ d \phi_i \)^2 + \frac{1}{6} \sum_{i=1,2} \( d\theta_i^2 
+ \sin^2 \theta_i ~ d \phi_i^2 \).
\label{t11}
\eeq
The base of the conifold is determined by
\beq
z_1 z_2 = z_3 z_4 , \quad z_i \in \Cbf,
\label{zdf}
\eeq
where by definition, $z_i \sim r^{3/2}$.  The angular
dependence of the $z_i$ that give rise to
\myref{t11} can be found in \cite{Klebanov:1998hh}.
We work in the near-horizon limit $r \ll L$
and approximate the ``warp factor'' $H(r) \approx (L/r)^4$.  
Unless otherwise stated, we will work in $L=1$ units (i.e., dimensions
are restored in what follows via $r \to Lr$, $x^\mu \to Lx^\mu$, etc.).

\subsection{Scalar fluctuations of the D7 brane embedding}
Following Levi and Ouyang \cite{Levi:2005hh},
the embedding of the D7 brane is characterized by
\beq
z_1 = r^{3/2} \sin \frac{\theta_1}{2}\sin \frac{\theta_2}{2}~
e^{\frac{i}{2}(\psi-\phi_1-\phi_2)} \equiv \mu > 0.
\eeq
Equivalently,
$r=r_0(\theta_i)$ and $\psi=\psi_0(\phi_i)$ with
\beq
r_0^{3/2} \sin \frac{\theta_1}{2} 
\sin \frac{\theta_2}{2}  = \mu , \quad
\psi_0 = \phi_1 + \phi_2 .
\label{loem}
\eeq
Thus the minimum radius is $r=\mu^{2/3}$.  Since the
$\adsv$ regime is in the near-horizon limit $r \ll 1$,
in order for the flavors on the D7 brane to behave
as fields in $\adsv$ in some regime it is necessary
that $\mu^{2/3} \ll 1$.  We assume that this
is true in what follows.  The embedding studied
here is among those that have been shown to
be consistent in the analysis of \cite{Arean:2004mm}.

The fluctuations of the brane in the two orthogonal
directions are given by the scalar modes\footnote{
This is the definition of $\chi$ that was actually
used in \cite{Levi:2005hh}, although version 1 of
the preprint had a typo \cite{LOpri}.}
\beq
r = r_0 (1 + \chi), \quad \psi = \psi_0 + 3 \eta,
\label{d7e}
\eeq
where $\chi,\eta$ are generally functions of
the eight worldvolume coordinates of the D7 brane.
The induced metric on the D7 brane is
\beq
(g_0)_{ab} &=& \diag( r_0^2 \eta_{\mu \nu}, g_{\theta_i \theta_j},
g_{\phi_i \phi_j} ), \nnn
g_{\theta_i \theta_j} &=&
\begin{pmatrix} \frac{1}{6} + \frac{1}{9} \cot^2 \frac{\theta_1}{2} &
\frac{1}{9} \cot \frac{\theta_1}{2} \cot \frac{\theta_2}{2} \cr
\frac{1}{9} \cot \frac{\theta_1}{2} \cot \frac{\theta_2}{2} &
\frac{1}{6} + \frac{1}{9} \cot^2 \frac{\theta_2}{2} \end{pmatrix}, \nnn
g_{\phi_i \phi_j} &=&
\begin{pmatrix} \frac{1}{6} \sin^2 \theta_1 + \frac{1}{9} (1+\cos \theta_1)^2 &
\frac{1}{9} (1 + \cos \theta_1)(1+ \cos \theta_2) \cr
\frac{1}{9} (1 + \cos \theta_1)(1+ \cos \theta_2) &
\frac{1}{6} \sin^2 \theta_2 + \frac{1}{9} (1+\cos \theta_2)^2 \end{pmatrix},
\eeq
where ``diag'' denotes a block diagonal matrix.
The Dirac-Born-Infeld (DBI) action,
\beq
S_{\text{DBI}} = - \tau_7 \int d^4x\,d^2\theta\,d^2\phi ~
\sqrt{ \varphi^\star(g) + \varphi^\star(B) + 2\pi \alpha' F },
\label{dbi}
\eeq
with $\tau_7 = (2\pi)^{-7} \alpha'^4 g_s^{-1}$, 
contains the scalar fluctuations \myref{d7e} through
perturbations of the pullback $\varphi^\star(g) =
g_0 + \delta g_0(\chi,\eta)$ of the 10d metric
to the D7 brane worldvolume.  We will return to the
worldvolume 2-form $F=dA$ in \S\ref{ggef} below.
The pullback of the NS-NS 2-form $\varphi^\star(B)$
will be neglected in our analysis, since we are
interested in fields that carry flavor quantum numbers.
Note that the B-field vanishes in the KW background.
To quadratic order, the action for the scalar fluctuations 
\myref{d7e} is:
\beq
S & = & -\tau_7 \int d^4x\,d^2\theta\,d^2\phi~ \bigg\{ \sqrt{-g_0}
\bigg[ \frac{g_0^{ab}}{2C} \( \p_a \chi \p_b \chi + \p_a \eta \p_b \eta \) \nnn
&& \quad + \frac{4}{C} (\sin^2 \frac{\theta_i}{2} )^{-1} \chi \p_{\phi_i} \eta
- \frac{2}{C^2} ( \sin^2\frac{\theta_i}{2} )^{-1}
\cot\frac{\theta_j}{2}\partial_{\theta_j}
(\chi \p_{\phi_i} \eta) \bigg] \nnn
&& \quad - \p_{\theta_i} \[ \frac{\sqrt{-g_0}}{C} \cot \frac{\theta_i}{2}
\( 3 \chi^2 + 2 \chi \) \] \bigg\},
\label{fula}
\eeq
where it has been convenient to define
\beq
C=1+ \frac{2}{3}\cot^2\frac{\theta_1}{2} +\frac{2}{3}
\cot^2\frac{\theta_2}{2},
\label{Cdf}
\eeq
and implicit sums over $a,b \in \{x^\mu,\theta_i,\phi_i \}$, with
$\mu \in \{ 0,\ldots,3\}$, and $i,j \in \{1,2\}$.
Thus we agree with eq.~(29) of Levi and Ouyang \cite{Levi:2005hh};
note, however, that we have explicitly included the
boundary terms that occur in the
simplifications (integration by parts) that lead to \myref{fula}.

Our interest is in the lightest states, which are not excited modes
on the internal compact space $T^{1,1}$.
Setting $\p_{\phi_i}=0$ and integrating over the $\phii$
coordinates we find:
\beq
S &=& -4 \pi^2 \tau_7 \int d^4 x\, d^2 \theta \bigg( \sqrt{-g_0}
\[ \frac{g_0^{mn}}{2C}  \( \p_m \chi \p_n \chi
+ \p_m \eta \p_n \eta \) \] \nnn
&& 
\quad - \p_{\theta_i} \[ \frac{\sqrt{-g_0}}{C} \cot \frac{\theta_i}{2}
\( 3 \chi^2 + 2 \chi \) \] \bigg),
\label{opp}
\eeq
where we now have indices $m,n \in \{x^\mu,\theta_i\}$.
To proceed further, we must extract the radial dependence that
has been hidden in the angular variables~$\theta_i$;
cf.~\myref{loem}.

\subsection{A radial reparameterization}
It is convenient to introduce a scaled radius:  $r_0 = \mu^{2/3} \rhat$.
We can express the D7 brane embedding \myref{loem} in the equivalent form
\beq
2 = \rhat^{3/2} ( \cos \tm - \cos \tp), \quad
\theta_\pm \equiv \half (\theta_1 \pm \theta_2).
\label{mpc}
\eeq
We will eliminate $\tp$ in favor of the coordinates $\rhat,\tm$.
The domain of $\tm$ depends on $\rhat$, and is
given by
\beq
\tm \in [-\theta_0(\rhat),\theta_0(\rhat)], \quad
\theta_0(\rhat) \equiv \frac{\pi}{2} - \sin^{-1} \rhat^{-3/2}.
\label{tmd}
\eeq
Taking into account the Jacobian of the transformation,
we have
\beq
&& \int_0^\pi d\theta_1 \int_0^\pi d\theta_2 (\cdots) =
\int_{1}^\infty d\rhat \int_{-\theta_0(\rhat)}^{\theta_0(\rhat)}
d \tm \frac{6}{\rhat^{5/2} \sin \tp(\rhat,\tm)} (\cdots),
\label{poiu}
\eeq
where
\beq
&& \sin \tp(\rhat,\tm) = \[ 1- ( \cos \tm - 2\rhat^{-3/2} )^2 \]^{1/2}.
\label{poit}
\eeq
We will reduce to a 5d effective theory by imposing
$\tm$ independence, corresponding to no excitation in
this compact coordinate.  The validity of this
for the eight-dimensional (8d) scalars $\chi,\eta$ rests on the fact that at fixed
$\rhat$ the D7 brane wraps a 3-manifold that is topologically
equivalent to $S^3$, as shown in the Appendix. 
In the reduction to 5d, the 8d scalars should be expanded on scalar harmonics
of this 3-manifold, which will include the constant
``$\ell=0$'' mode.  This translates into a $\tm,\phii$ independent
mode in the coordinates that are used here.

Note that we retain the parameterization \myref{d7e}
of the D7 embedding fluctuations, although we have
now taken $r$ as a parameter of the D7 brane worldvolume.
In fact, it is not difficult to show that
$\chi$ can be reinterpreted as a
fluctuation of $\theta_+$ away from $\theta_+=\theta_+(\rhat,
\theta_-)$ determined from~\myref{mpc}.

The quantity $C$ that appears in \myref{Cdf} can
be written as:
\beq
C = \third \[ -1  + 4\rhat^{3/2} \cos \tm
+ \rhat^3 (1 - \cos 2\tm) \].
\eeq
We also find that the metric density of the
old coordinates takes the form
\beq
\sqrt{-g_0} \equiv \mu^{8/3} \sqrt{-g(\rhat,\tm)}, \quad
\sqrt{-g(\rhat,\tm)} = 
\frac{C}{9} \rhat ( \rhat^{3/2} \cos\tm - 1).
\label{scom}
\eeq
For the metric $\hat g_{mn}$ in the new coordinates 
$m,n \in \{x^\mu,\rhat,\tm,\phi_i\}$, we have
\beq
\sqrt{- \ghat} = \frac{6 \mu^{8/3} \sqrt{-g(\rhat,\tm)}}
{\rhat^{5/2} \sin \tp(\rhat,\tm)} =
\frac{2 \mu^{8/3} C ( \rhat^{3/2} \cos\tm - 1)}
{3 \rhat^{3/2} \sin \tp(\rhat,\tm)},
\label{nmt}
\eeq
in accordance with \myref{poiu}.  The components of
the new metric $\ghat$ are given in \myref{nmet} below.
In the new coordinates we obtain the action:
\beq
S &=& -4 \pi^2 \tau_7 \int d^4 x \int_{1}^\infty d\rhat 
\int_{-\theta_0(\rhat)}^{\theta_0(\rhat)} d \tm  \bigg\{
\sqrt{-\ghat} \nnn
&& \times \[ \frac{\hat g^{ij}(\rhat,\tm)}{2C(\rhat,\tm)} 
\( \p_i \chi \p_j \chi + \p_i \eta \p_j \eta \) \]
+ {\rm t.d.} \bigg\},
\label{rtme}
\eeq
where ``t.d.'' = total derivatives,
$\rhat,\tm$ dependence has been made explicit,
and now $i,j \in \{ \mu, \rhat \}$.
The transformed inverse metric $\hat g^{ij}$ is just
the one that follows from the change of coordinates.
We only need:\footnote{There are also $\hat g^{\rhat \tm}$
and $\hat g^{\phi_i \phi_j}$ components that we
are able to ignore because of our angular independence
assumption.  These components of course make
an implicit appearance in the overall measure that appears
in \myref{rtme}.  Cf.~\myref{nmet} below.}
\beq
\hat g^{ij} = 
\begin{pmatrix}
\hat g^{\mu \nu} & 0 \cr 0 & \hat g^{\rhat \rhat}
\end{pmatrix} =
\begin{pmatrix}
\mu^{-4/3} \rhat^{-2} \eta^{\mu \nu} & 0 \cr 0 & (1-C^{-1})\rhat^2
\end{pmatrix}.
\label{ghat}
\eeq
We still must integrate over
the $\tm$ dependence that appears explicitly in the Lagrangian.
To this end we define the following $\rhat$-dependent 
functions:\footnote{Note that these definitions are
expressed in terms of
the old metric density with $\mu^{8/3}$ scaled out, the
quantity $\sqrt{-g(\rhat,\tm)}$
defined in \myref{scom}, rather than the new metric density \myref{nmt}.}
\beq
&& F_1(\rhat) \equiv \int_{-\theta_0(\rhat)}^{\theta_0(\rhat)} d \tm 
\frac{ \sqrt{-g} }{\sin \tp C} (\rhat,\tm), \nnn
&& \tilde F_1(\rhat) \equiv \int_{-\theta_0(\rhat)}^{\theta_0(\rhat)} d \tm 
\frac{ \sqrt{-g} (C-1)}{\sin \tp C^2}  (\rhat,\tm).
\label{f12}
\eeq
Then the effective 5d action for $\tm,\phi_i$ independent
modes is:
\beq
S &=& -24 \pi^2 \mu^{8/3} \tau_7 \int d^4 x \int_{1}^\infty d\rhat
\bigg\{ \half \mu^{-4/3} \rhat^{-9/2} F_1(\rhat)
\eta^{\mu \nu} \( \p_\mu \chi \p_\nu \chi + \p_\mu \eta \p_\nu \eta \) \nnn
&& + \half \rhat^{-1/2} \tilde F_1(\rhat) [ (\p_\rhat \chi)^2
+ (\p_\rhat \eta)^2 ] + {\rm t.d.} \bigg\} .
\label{f12a}
\eeq

\begin{figure}
\begin{center}
\includegraphics[width=2.7in,height=3.5in,angle=90]{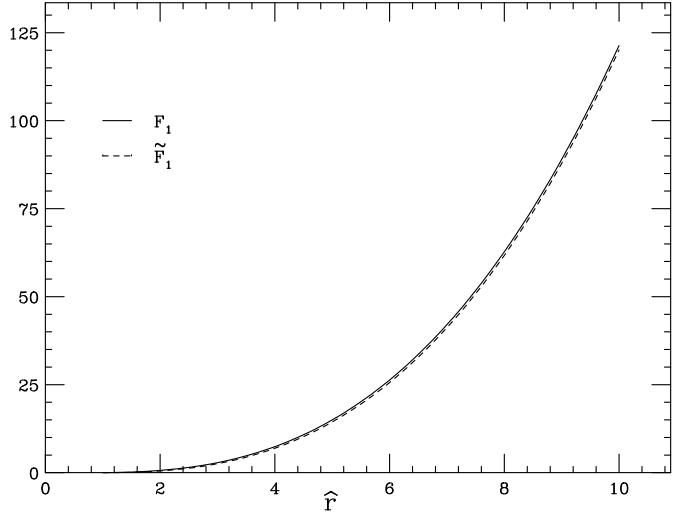}
\caption{The functions $F_1(\rhat)$ and $\tilde F_1(\rhat)$,
defined in \myref{f12}.  Both functions are well-described
by the approximation \myref{f1ex}.
\label{F1fig}}
\end{center}
\end{figure}

In Fig.~\ref{F1fig} we show the
functions $F_1$ and $\tilde F_1$,
each of which vanishes at $\rhat=1$.
We will have more to say about these functions below.

\subsection{Boundary terms}
\begin{figure}
\begin{center}
\includegraphics[width=4.5in,height=2.8in,bb=100 400 500 670,clip]{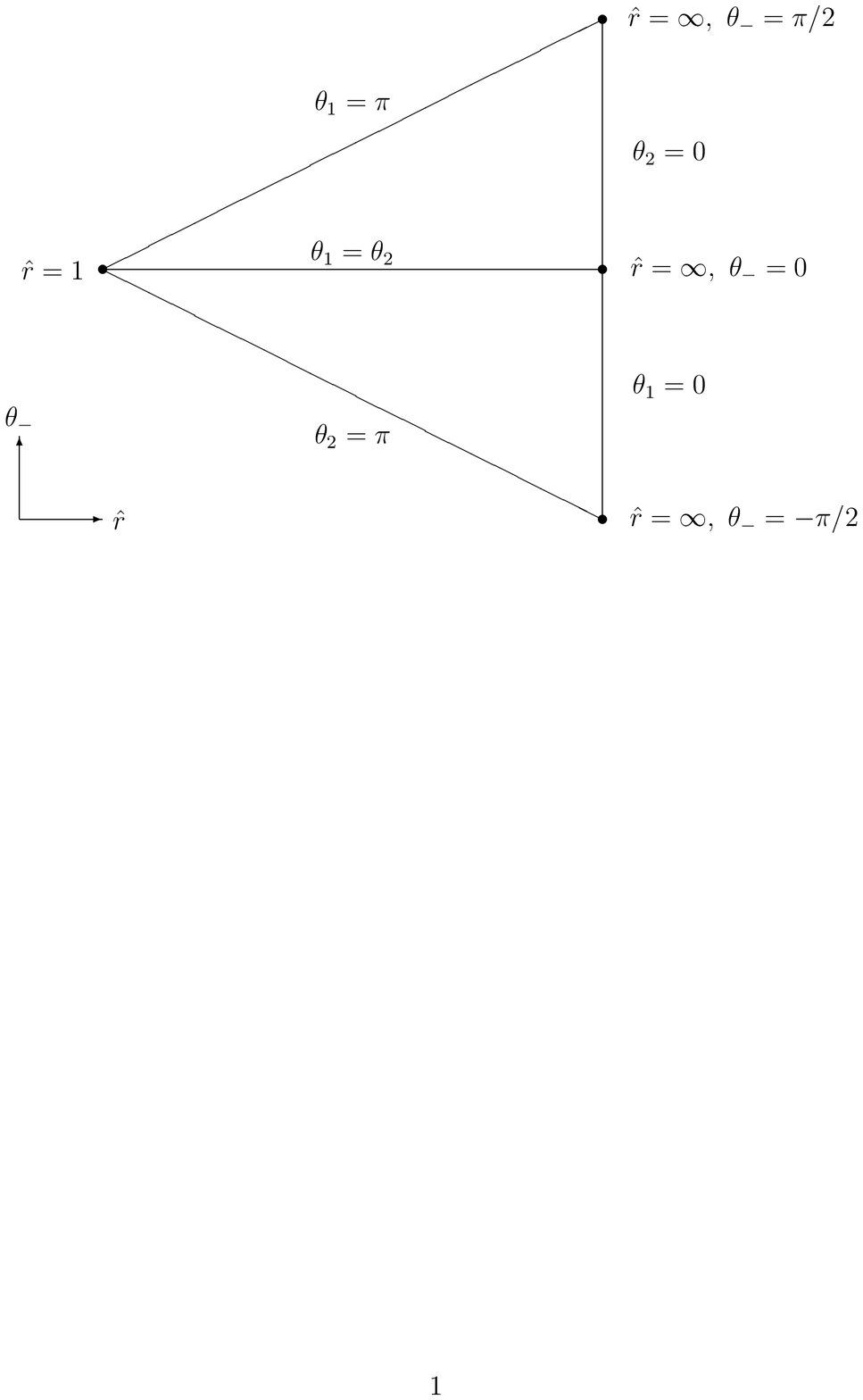}
\caption{Relation between coordinates $\theta_{1,2}$ and
$\rhat,\theta_-$, taking into account the $\rhat$-dependent
domain of $\theta_-$ for the D7 embedding, given
in \myref{tmd}.  The boundaries at $\theta_{1,2} = \pi$
correspond to $\theta_- = \pm \theta_0(\rhat)$,
where $\theta_0(\rhat)$ is monotonically increasing
in $\rhat$, from $\theta_0(1)=0$ to $\theta_0(\infty)= \pi/2$.
The boundary at $\rhat = \infty$ consists of two
segments, distinguished by positive and negative $\theta_-$.
These correspond to $\theta_{2,1}=0$.  The $\rhat=1$
boundary is just a point in this 2d subspace. \label{fdf}}
\end{center}
\end{figure}

Next, we briefly describe the total derivative terms
that appear in \myref{opp}.  These yield boundary
terms at $\theta_{1,2} = 0$ and $\pi$.  
To elucidate
the relation of these boundaries to the $\rhat,\tm$
coordinates, we have mapped $\rhat$ into a finite domain
in Fig.~\ref{fdf}.  This figure takes into account
the $\rhat$-dependent domain of $\tm$, Eq.~\myref{tmd}.  Along the
$\theta_1=\pi$ boundary, we can write $\theta_2=\theta_2(\rhat)
=\pi-2\theta_0(\rhat)$.
A similar statement holds along the $\theta_2=\pi$ boundary,
with $\theta_1=\pi-2\theta_0(\rhat)$.
It is therefore straightforward to express these two
boundary terms as integrals over $\rhat$, with
$\tm = \pm \theta_0(\rhat)$ (it will be seen below
that these boundary terms vanish.).
At the $\theta_2=0$
boundary, we can write $\theta_1=2 \tm$.
A similar statement holds on the $\theta_1=0$
boundary, where $\theta_2= -2\tm$.
These two boundary terms can therefore
be expressed as integrals over $\tm \in [0,\pm \pi/2]$
with $\rhat \to \infty$.  
The latter limit will require some care.
Corresponding to the total derivative
terms in \myref{opp}, we define
\beq
H_i^{(1)}(\theta_1,\theta_2) &=& 2 \sqrt{-g_0} ~ C^{-1}  \cot \frac{\theta_i}{2} ~ \chi,
\nnn
H_i^{(2)}(\theta_1,\theta_2) &=& 3 \sqrt{-g_0} ~ C^{-1}  \cot \frac{\theta_i}{2} ~ \chi^2.
\eeq
Then it is easy to show that the total derivatives
can formally be written as the following boundary action:
\beq
&& S_b = -8 \pi^2 \tau_7 \int d^4 x \sum_{i,\alpha=1,2} 
\bigg\{ \int_1^\infty d\rhat ~ \left|
\frac{\p \theta_0}{\p \rhat} \right| \[ H_i^{(\alpha)}(\pi,\pi-2\theta_0)
+ H_i^{(\alpha)}(\pi-2\theta_0,\pi)\] \nnn
&& \quad + \; \int_0^{\pi/2} d\tm ~ H_i^{(\alpha)}(2\tm,0)
+ \int_{-\pi/2}^0 d\tm ~ H_i^{(\alpha)}(0,-2\tm) \bigg\} \nnn
&=& -8 \pi^2 \tau_7 \int d^4 x \sum_{i,\alpha=1,2} \bigg\{ 
\int_0^{\pi/2} d\tm ~ H_i^{(\alpha)}(2\tm,0)
+ \int_{-\pi/2}^0 d\tm ~ H_i^{(\alpha)}(0,-2\tm) \bigg\}.
\eeq
Here we provide an intermediate
expression in order to emphasize that the boundaries 
with $\tm=\pm \theta_0$ (first line)
give vanishing contributions.  This is fortunate, since
they would otherwise give bulk contributions
as far as the 5d reduction is concerned.
The final expression is just the $\rhat \to \infty$
boundary terms, which should be interpreted in terms of limits.
Boundary conditions on $\chi$ will have
to be imposed such that the result is well-defined.

\subsection{Gauge fields}
\label{ggef}
Under dimensional reduction, the 8d vector
boson $A_a$ of the worldvolume D7 brane $U(1)$ gauge theory
decomposes into a 5d gauge field $A_\mu$ 
and three real scalars $A_{\tm,\phii}$.  We now extract the quadratic
action and equations of motion for these fields.  We will
then make some brief comments regarding these modes.
We will point out the difficulties that arise
from the vector harmonic analysis of the $A_{\tm,\phii}$ modes
due to the nontrivial 3-manifold that they
are compactified on.  

It is straightforward to expand the DBI action \myref{dbi}
to quadratic order in the 8d field strength $F_{ab}$.
In addition, the Wess-Zumino (WZ) term needs to
be considered, due to the nontrivial 4-form background
that is present in the KW construction:
\beq
S_{\text{WZ}} = \half (2\pi\alpha')^2 \tau_7 
\int C_4 \wedge F \wedge F, \quad
C_4 = \mu^{8/3} \rhat^4 dx^0 \wedge \cdots \wedge dx^3.
\eeq
The $F \wedge F$ that appears in the WZ 
action can only have ``legs'' along the
$\rhat,\tm,\phii$ directions, which we will
label collectively by $\alpha,\beta$ etc.  Altogether,
we have:
\beq
S_{F^2} = (2\pi\alpha')^2 \tau_7 \int d^4x\,d \rhat\,d \tm\, d^2 \phi\,
\bigg\{ -\fourth \sqrt{-\ghat} F_{ab} F^{ab} 
+ \frac{1}{8}\mu^{8/3}\rhat^4 \e^{\alpha \beta \gamma \delta}
F_{\alpha \beta} F_{\gamma \delta} \bigg\}~,
\eeq
where the metric is given by:
\beq
\label{nmet}
&& \ghat^{\mu \nu} = \mu^{-4/3} \rhat^{-2} \eta^{\mu \nu}, \quad
\ghat^{\rhat \rhat} = (1-1/C) \rhat^2, \nnn
&& \ghat^{\rhat \tm} = \rhat^{5/2} C^{-1} \sin \tm, \quad
\ghat^{\tm \tm} = (4 \rhat^{3/2} \cos \tm + 3C - 1)/2C, \\
&& \ghat^{\phii \phij} = \frac{\rhat^3}{4C} \begin{pmatrix} 
\frac{5-\cos(\tm - \tp)}{\cos^2 \frac{1}{2}(\tm + \tp)} & -4 \cr
-4 & \frac{5- \cos(\tm + \tp)}{\cos^2\frac{1}{2}(\tm - \tp)} \end{pmatrix}~,
\nonumber
\eeq
with $\theta_\pm$ defined in \myref{mpc},
and $\sqrt{-\ghat}$, given in \myref{nmt}, contributing
the same coefficient $\mu^{8/3}$ as appears in the WZ term.
The equations of motion are
\beq
0 = \p_a (\sqrt{-\ghat} F^{ab}) - 4 \mu^{8/3} \rhat^3 \e^{bjk} \p_j A_k,
\eeq
where $\e^{bjk} = 0$ unless $b \in \{ \tm, \phii \}$, and
by definition $j,k \in \{ \tm, \phii \}$, the coordinates
of the internal 3-manifold $X_3$ at fixed radius $\rhat$ (see the Appendix).

Since the fields $A_k$ are vectors on $X_3$, it is
necessary to expand them on vector harmonics of $X_3$.
This analysis is far from trivial, for a couple of reasons.
First, the metric of $X_3$ depends on $\rhat$.  Second,
even at $\rhat \to \infty$, where the metric of $X_3$ becomes
independent of $\rhat$, the geometry of the 3-manifold
is not simple, as is discussed in the Appendix.  
To determine the vector harmonics requires an analysis comparable to 
that done in 
Refs.~\cite{Gubser:1998vd,Jatkar:1999zk,Ceresole:1999zs,Ceresole:1999ht}
for the 5-manifold $T^{1,1}$.

It is easy to check that the 5d vector boson $(A_\mu,A_\rhat)$
has vanishing bulk mass.  Note that this mode corresponds
to the constant scalar harmonic on the compact 3-manifold $X_3$.
From the $\adsv$ supersymmetry that is present in the model,
we know that there must be a 5d real scalar partner with bulk
mass-squared $m^2 = -4/L^2$.  This must emerge from the analysis
of the modes $A_\tm,A_\phii$, and would be a nontrivial check
of the supersymmetry that is beyond the scope of the present 
work. In addition there will also be a massless (Dirac) fermion 
corresponding to the gaugino.

\mys{Relation to 5d effective theories}
\label{r5d}
In this section we relate the above string
construction to the sort of 5d effective
theories that are generally contemplated
in phenomenological applications~\cite{Gherghetta:2006ha}.  
Our first task is to show that, in an appropriate limit
the scalars $\chi,\eta$ described above behave like fields
in $\adsv$.

\subsection{The $\adsv$ regime}
\label{adr}
The action for a massive real scalar field in
a semi-infinite slice of $\adsv$ is given by
\beq
S = -\half \int d^4x \int_R^\infty dr\,\bigg\{ \frac{r}{L}
\eta^{\mu \nu} \p_\mu \phi \p_\nu \phi + \frac{r^5}{L^5}
\p_r \phi \p_r \phi + \frac{r^3}{L^3} m^2 \phi^2 \bigg\},
\label{can5}
\eeq
where our coordinate conventions are summarized by the metric:
\beq
ds_5^2 = \frac{r^2}{L^2} \eta_{\mu\nu}dx^\mu dx^\nu
+ \frac{L^2}{r^2} dr^2.
\eeq
We would like to find a regime where the 
action \myref{can5} is a good approximation to
the action of $\chi,\eta$ \myref{f12a}.
We expect this to be possible since
in the near-horizon regime ($r \ll L)$ 
the supergravity background is
$\adsv \times T^{1,1}$.  The complication
is that the D7 brane is embedded into this space
in a way that constrains angles in $T^{1,1}$
to be related to the $\adsv$ radius $r$ via
\myref{loem}.  This complicates the
radial dependence of the $\chi,\eta$ action,
as can be seen from the various expressions in 
the previous sections.

However, from \myref{mpc} it is easy to
see that in the $\rhat \to \infty$ limit,
the embedding approaches $\tm \equiv \tp$,
which is independent of the radius $\rhat$.
Recall that since $\rhat =  r / \mu^{2/3}$, 
$\rhat \gg 1$ corresponds to $r \gg \mu^{2/3}$ and
therefore to remain in the near-horizon limit, we
also require $r \ll L$.  Thus we want to
examine the above expressions in the regime
$\mu^{2/3} \ll r \ll L$.  In the $L=1$ units
used above, this is equivalent to
\beq
\rhat \gg 1, \quad \mu^{2/3} \ll 1~.
\eeq
We will extract the leading order Lagrangian
under these assumptions and compare to \myref{can5}.
Our finding is that the usual, conformally coupled 
scalar action is recovered.  Deviations from this
action due to subleading terms (logarithmic in $\rhat$) are related to
the breakdown of conformal invariance near the
mass threshold of the flavors of the dual gauge
theory, corresponding to the end of the D7 brane probe
at $r=\mu^{2/3}$.

After a careful numeric and analytical 
study of the integrals \myref{f12}, we find that
\beq
F_1 \approx \frac{1}{6} \rhat^{5/2} \ln \rhat~,
\label{f1ex}
\eeq
to an approximation that is good to
five significant digits at all values of
$\rhat$.  It is possible to obtain an
exact expression for $F_1$ in terms of
elliptic functions, which gives \myref{f1ex},
corrected by subleading logs.
Also, $\tilde F_1 \approx F_1$ in an
approximation that becomes exact in
the $\rhat \to \infty$ limit; in fact, it can
be seen from Fig.~\ref{F1fig} that the
two functions are nearly equal for all
values of $\rhat$.  However, for $\rhat = \ord{1}$,
in relative terms the r.h.s.~of \myref{f1ex} is a poor approximation
to $\tilde F_1$.

To obtain an action corresponding to the
form \myref{can5}, the following
field redefinitions must be made:
\beq
\chi = \rhat^{3/2} (\ln \rhat)^{-1/2} \chi', \quad
\eta = \rhat^{3/2} (\ln \rhat)^{-1/2} \eta' .
\label{cerd}
\eeq
Note that to an excellent approximation, this is just a
rescaling by $F_1^{-1/2}$ and an appropriate power of $\rhat$.
Taking into account the powers of $\rhat$
that arise from \myref{f12a} and \myref{f1ex}, the following
radial gradient term occurs in the Lagrangian:
\beq
\rhat^2 \ln \rhat (\p_\rhat \chi)^2 &=& \rhat^5 (\p_\rhat \chi')^2
- \( \frac{15}{4} - \frac{1}{2\ln \rhat} + \frac{1}{4 (\ln \rhat)^2}  \)
\rhat^3 \chi'^2 + \text{t.d.},
\label{yuyu}
\eeq
with an identical equation for $\eta$.
Substitution of \myref{cerd}-\myref{yuyu} into \myref{f12a}
yields the bulk action 
\beq
S(\chi') &\approx& -2 \pi^2 \mu^{8/3} \tau_7 \int d^4x \int_{\hat R}^\infty
d\rhat \bigg\{ \frac{\rhat}{\mu^{4/3}} \eta^{\mu \nu} \p_\mu \chi' \p_\nu \chi'
\nnn && \quad
+ f(\rhat) \[ \rhat^5 (\p_\rhat \chi')^2 + \rhat^3 m^2(\rhat) \chi'^2 \] \bigg\},
\label{adeb}
\eeq
and an action for $\eta'$ that is the same.  We have
introduced the ratio 
\beq
f(\rhat) = \tilde F_1(\rhat)/F_1(\rhat) \approx 1 \quad
\text{for} \quad \rhat \gg 1.
\label{oqq}
\eeq
In \myref{adeb}, a cutoff $\hat R \gg 1$ on the radial integration
has been introduced.  The effects of integration
over $\rhat \in [1,\hat R]$, as well as total derivative terms
could, for instance, be incorporated into an effective boundary action.
This is discussed in \S\ref{eba} below.  For $\hat R \gg 1$,
it is a good approximation to take $f(\rhat)=1$ in \myref{adeb},
which is what we do in the following.

\begin{figure}
\begin{center}
\includegraphics[width=2.7in,height=3.5in,angle=90]{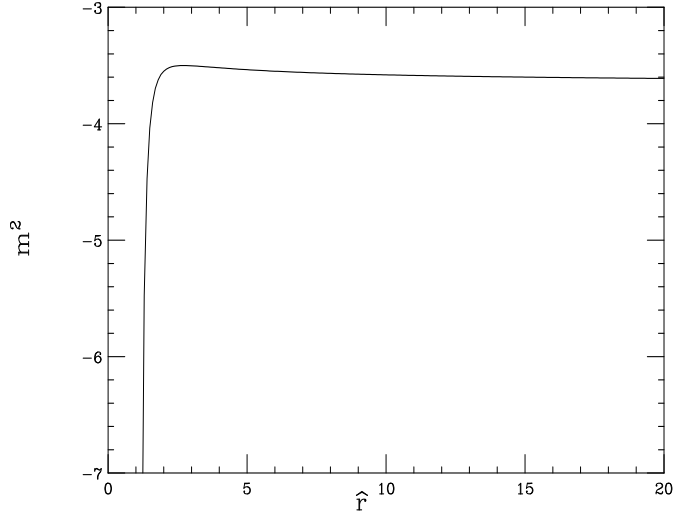}
\caption{The mass that is defined in \myref{adec}.
It approaches the constant value $-15/4$ at
large $\rhat$.  The infinitely negative value
at $\rhat \to 1$ gives rise to the Dirichlet
boundary conditions \myref{dbc}.
\label{msq}}
\end{center}
\end{figure}

The $\rhat$-dependent mass is:
\beq
m^2(\rhat) &=& - \frac{15}{4} + \frac{1}
{2 \ln \rhat} - \frac{1}{4 (\ln \rhat)^2}.
\label{adec}
\eeq
The terms proportional to $1/\ln \rhat$ or its square
are subleading in the $\rhat \gg 1$ regime.
In Fig.~\ref{msq} we display the mass \myref{adec}.
In the $\rhat \to 1$ limit, $m^2(\rhat)$ becomes infinitely negative.  
As a consequence,
physical solutions must satisfy Dirichlet boundary
conditions:
\beq
\lim_{\rhat \to 1} \chi'(\rhat) = \lim_{\rhat \to 1} \eta'(\rhat) = 0.
\label{dbc}
\eeq
The approach to zero at $\rhat=1$ must be stronger
than $\ln \rhat$.  In ref.~\cite{Levi:2005hh},
it is stated that for regularity the original fields $\chi,\eta$
should satisfy Neumann boundary conditions at $\theta_1=
\theta_2=1$, equivalent to $\rhat=1$.  Taking into account
the factor $\sqrt{\ln \rhat}$ that appears in \myref{cerd},
it is clear that a finite $\chi,\eta$ at $\rhat=1$ implies
a vanishing $\chi',\eta'$.  Thus the two findings
on boundary conditions at $\rhat=1$ are consistent.

The mass degeneracy for $\chi',\eta'$
is a consequence of 5d supersymmetry
in the $\adsv$ regime:  the real scalars
$\chi',\eta'$ must combine into a complex scalar
of a 5d supersymmetry hypermultiplet.

If we return to the variables $r=\rhat \mu^{2/3}$ and 
$R=\hat R \mu^{2/3}$, and reintroduce the $\adsv$ radius $L$ explicitly
by scaling the fields $\chi' \to \chi'/L$, $\eta' \to
\eta'/L$, we obtain
\beq
S(\chi') \approx
-2 \pi^2 L^{-5} \tau_7 \int d^4x \int_{R}^\infty
dr\, \bigg\{ \frac{r}{L} \eta^{\mu \nu} \p_\mu \chi' \p_\nu \chi'
+ \frac{r^5}{L^5} (\p_r \chi')^2 
- \frac{15}{4L^2} \frac{r^3}{L^3} \chi'^2 \bigg\},
\eeq
and similarly for $\eta'$.  Here we have neglected
the subleading logs in the mass terms and the prefactor \myref{oqq}.

For $\chi',\eta'$ we obtain a negative mass-squared:
\beq
m^2 = - \frac{15}{4L^2} > - \frac{4}{L^2},
\eeq
where the inequality expresses the fact that the
mass-squared satisfies the Breitenlohner-Freedman
bound \cite{Breitenlohner:1982bm,Breitenlohner:1982jf}.
The explanation of the negative mass-squared is similar
to that given for probe branes in $\adsv \times S^5$ geometries
\cite{Karch:2000gx,Karch:2002sh,Kruczenski:2003be}.
We show in the Appendix that
at fixed $\rhat$ the D7 brane wraps a 3-manifold that
is topologically equivalent to $S^3$.
The radius of this $S^3$ shrinks to zero
as $\rhat \to 1$.  It is therefore a
topologically trivial 3-cycle in the conifold.
The negative mass-squared corresponds to a ``slipping
mode.''

The bulk mass is independent of the scale $\mu^{2/3}$,
as it should be, since that is an IR boundary scale.
The $\mu$ dependence has also disappeared from
the overall factor in front of the action,
which also makes sense from this perspective.
The purely numerical value $-15/4$ arises from the radial
independence of the embedding $\tm \equiv \tp$
that occurs in the $\rhat \to \infty$ limit.
In fact, $-15/4$ is interesting because it is nothing but the
conformally coupled scalar in $\adsv$ \cite{Kim:1985ez,Gunaydin:1984fk}, 
and corresponds to the Laplacian eigenvalue for
the lowest mode of scalar harmonics on $T^{1,1}$ 
\cite{Gubser:1998vd,Jatkar:1999zk,Ceresole:1999zs,Ceresole:1999ht}.
Since we are considering only modes that
are independent of the $T^{1,1}$ coordinates, they
do not ``know'' that the D7 brane is actually restricted
to a submanifold of $T^{1,1}$.  This explains
the equivalence to the lowest $T^{1,1}$ scalar harmonic.

The $\ord{1/\ln \rhat}$ mass terms in \myref{adec}
represent the leading effect of the breaking of
conformal symmetry due to $\mu \not= 0$.  In the
dual gauge theory this parameter is related to
the Yukawa and mass parameters of massive ``flavor
probes'' that have been added to the original KW construction.  At scales
where the mass of these flavors is noticeable,
the conformal symmetry is broken.  The threshold
for these flavors corresponds to $\rhat=1$, where
the D7 brane ends.  Far away from this tip, at
$\rhat \gg 1$, the dual gauge theory is at energy
scales far above the threshold, where universal
behavior dominates and scaling dimensions become apparent.

\subsection{Effective boundary action}
\label{eba}
On physical grounds, there is one boundary condition (BC)
at $\rhat=1$ that must be satisfied:  since the fields
end there, and this should happen continuously, we have
Dirichlet BCs~\myref{dbc}.  It was seen above that
this naturally emerges from infinitely negative
mass terms.  Solving the equations of motion
in the region $\rhat \in \{ 1,\Rhat \}$, we can impose
one more BC, generally Cauchy, at $\Rhat$.  Thus,
we obtain a discrete set of permissible Cauchy
BCs at $\Rhat$, parameterized by the functional
condition:
\beq
G[\chi'(x,\Rhat),\p_\rhat \chi'(x,\Rhat)] = 0 \quad
\forall x,
\eeq
and similarly for $\eta'$.  This constraint may
then be translated into an effective boundary action
involving a Lagrange multiplier $\psi$:
\beq
S'=\int d^4x ~ \psi(x) G[\chi'(x,\Rhat),\p_\rhat \chi'(x,\Rhat)].
\eeq
$\psi$ is interpreted as a boundary field; we
can give it dynamics on the boundary, provided
it still has the effect of setting $G = 0$ to a good
approximation.

From Fig.~\ref{msq}, we see that to a first
approximation the $\rhat \approx 1$ effects just
impose Dirichlet BCs at $\rhat = \Rhat = \ord{1}$:
\beq
G[\chi'(x,\Rhat),\p_\rhat \chi'(x,\Rhat)] \approx \chi'(x,\Rhat).
\eeq
That is, Fig.~\ref{msq} shows that $\Rhat$ of just ``a few''
suffices to approach the constant value of $m^2 = -15/4$;
the requirement $\Rhat \gg \rhat$ is stronger than is
actually needed, in order to render the log corrections in \myref{adec}
negligible.  Thus, the leading order behavior is just that
of a conformally coupled scalar with Dirichlet BCs
at the IR boundary.

\subsection{Auxiliary scalar action}
\label{asa}
Here we briefly touch on an alternative
effective description of the small $\rhat$
behavior.  The approach here is modeled after what
was done in \cite{Brummer:2005sh}.
One introduces two scalars $h_{\chi',\eta'}$ 
to imitate the effect of the $\rhat$-dependent
part of the masses \myref{adec}.  These auxiliary
scalars are static, in the sense that for the
modes that couple to $\chi',\eta'$, we
can neglect dependence on 4d spacetime coordinates.
For this to work, it is necessary to replace
the $\rhat$-dependent parts of the mass terms for $\chi',\eta'$ with
\beq
&& -2 \pi^2 \mu^{8/3} \tau_7 \int d^4x \int_{\hat R}^\infty
d\rhat\, \bigg\{\rhat^5 \[ (\p_\rhat h_{\chi'})^2 + (\p_\rhat h_{\eta'})^2 \] \nnn
&&+\,\rhat^3 \[ V(h_{\chi'},h_{\eta'}) + h_{\chi'} \chi'^2
+ h_{\eta'} \eta'^2\] \bigg\} .
\eeq
The potential $V$ is engineered such that 
once the equations of motion for $h_{\chi',\eta'}$ are imposed
the profile of the auxiliary scalars is just
\beq
h_{\chi'} = h_{\eta'} = m^2(\rhat) + \frac{15}{4}.
\eeq
We will not pursue this effective description further,
since the ``microscopic'' description of \S\ref{eba} that is
available from the string construction is more
fundamental and elegant.  The only point that
we wish to make is that the unusual small $\rhat$
behavior of the scalar action can be mimicked by a coupling
to a quasi-static scalar with a nontrivial profile
for its lowest mode---something that a low-energy
phenomenologist might be more likely to consider.

\mys{Conclusions and outlook}
\label{conc}
The introduction of probe D7 branes in the Klebanov-Witten 
background provides a more fundamental description of 
5d phenomenological models in a slice of $\adsv$.
In this article we have concentrated on the
$\adsv$ regime that exists for a single D7 brane
embedded into the Klebanov-Witten background and derived
the effective 5d action for the scalar fluctuations. 
Whereas there is a significant departure from the
conventional scalar in $\adsv$ near the end
of the D7 brane, far away from that region the D7 embedding fluctuations
become conformally coupled scalars of $\adsv$. Furthermore by supersymmetry
there are also (massless) 5d bulk fermions.

In addition we have shown that in the $\adsv$ regime 
there are massless gauge fields. These fields mimic
the bulk gauge fields considered in 5d phenomenological 
models in a slice of $\adsv$. Again by supersymmetry 
we then infer that in the 5d bulk there are also 
massless (Dirac) fermions and a massive scalar with $m^2 =-4/L^2$.
Thus, probe D7 branes can provide all the necessary 5d bulk fields 
required for phenomenological model-building.

The simple setup that we have considered in this article 
can be generalized to provide a more realistic 
5d phenomenological model that incorporates the Standard Model, 
although a number of outstanding questions remain.

In particular one would like a full understanding of the
5d supergravity that occurs when the super-D7 brane
effective action and type IIB supergravity is
reduced on the 8d subspace $AdS_5 \times X_3$,
where we recall that $X_3$ is the $\rhat$-dependent D7 embedding 
into $T^{1,1}$.  At $\rhat \gg 1$, this would determine the complete 
multiplet structure for the bulk 5d supergravity and matter
fields.  Eventually
supersymmetry will also need to be broken 
so that a realistic low-energy spectrum is obtained. One possible
way would be to study flux compactifications as in GKP~\cite{Giddings:2001yu}.

To construct models with semi-realistic
gauge groups in the bulk, multiple D7 branes need to be 
considered---corresponding to a nonabelian gauge group
generalization.  Standard model matter can then be obtained 
by studying intersecting D7 brane models, where strings stretched 
between multiple D7 branes in the internal compact coordinates 
gives rise to matter with the usual Standard Model quantum numbers.

Normally 5d phenomenological models are compactified on $S^1/Z_2$ 
orbifolds, with corresponding bulk and boundary masses. Thus,
a detailed examination of the effective boundary action or 
auxiliary scalar action, as sketched in \S\ref{eba}-\ref{asa},
would be necessary. This may also require studying
the D7-brane fermion action for the Klebanov-Witten background, 
following the techniques of \cite{Bergshoeff:1996tu,Grana:2002tu,Marolf:2004jb}, 
supplemented by an analysis of spinor harmonics on $X_3$.
However, as noted earlier, information about the 
bulk fermion masses already follows from the scalar
mass-squared analysis, due to $\adsv$ supergravity constraints.

The most important aspect of the 5d phenomenological models is
their dual holographic interpretation as composite 4d 
theories~\cite{Gherghetta:2006ha}. The probe D7 branes introduce 
fundamental ``quarks'' in the dual gauge theory. Identifying the 
corresponding operators in the dual gauge theory,
especially with a realistic Standard Model spectrum, would elucidate
the holographic correspondence of composite states, like the top quark
and Higgs scalar field. This remains one of the most interesting 
avenues to study further.

Furthermore various refinements could also be introduced to the simple 
Klebanov-Witten construction that has been considered in this work.
As mentioned in the Introduction, one could introduce an IR cutoff
for non-probe modes by generalizing to the Klebanov-Strassler 
background \cite{Klebanov:2000hb}.
Here, the conifold \myref{zdf} is deformed:
\beq
z_1 z_2 - z_3 z_4 = \e^2.
\eeq
The parameter $\e$ determines the IR cutoff, and 
consistency of the supergravity theory requires a background
three-form flux.  This is a significant complication
for the spectral computation. The D7 brane probes of this background have
been studied, for instance, in \cite{Sakai:2003wu,Kuperstein:2004hy}.
The embeddings that were choosen are somewhat different
from \myref{loem}.  For all these D7 embeddings,
the main results will be essentially the same:
there is an $\adsv$ regime far away from where the
D7 brane ends; the end of the D7 brane can be replaced
by an effective boundary action, or an auxiliary
scalar; the small 5d radius regime, near
where the D7 brane ends, differs significantly
from $\adsv$.

In summary, probe D7 branes in the Klebanov-Witten background
provide a more fundamental description of 5d phenomenological models 
in a slice of AdS that solve the hierarchy problem. This framework
allows bulk fields to be introduced and leads to the possibility of 
explicitly constructing the dual theory.

\vspace{35pt}

\noindent
{\bf \Large Acknowledgements}

\vspace{5pt}

\noindent
We are grateful to Shamit Kachru for helpful remarks during the initial stages of 
this work. We thank Gianguido dall'Agata, Arthur Hebecker,
and Angel Uranga for useful discussions.
This work was supported in part by a Department of Energy grant 
DE-FG02-94ER40823 at the University of Minnesota. TG is also supported by
an award from the Research Corporation and acknowledges the hospitality of the
CERN Theory Division where this work was completed.

\myappendix

\mys{Geometric details of the D7 brane embedding}
\label{gede}
Here we provide some brief remarks on the
geometry of the D7 embedding relative to the
$AdS_5 \times T^{1,1}$ background.

First recall the standard argument that
shows that $T^{1,1}$ has isometry group 
$SU(2) \times SU(2) \times U(1)$.
The conifold equation \myref{zdf} that defines $T^{1,1}$ may
be expressed alternatively in coordinates
\beq
z_1 = w_1 + i w_2, \quad z_2 = w_1 - i w_2, \quad
z_3 = w_3 + i w_4, \quad z_4 = -(w_3 - i w_4),
\eeq
yielding
\beq
\sum_i w_i^2 = \det(w_4 1_2 + i \s_a w_a) = 0.
\label{zdf2}
\eeq
Here we have expressed the constraint in terms of
a complex quaternion equation, which has the
$SU(2)_1 \times SU(2)_2 \times U(1)$ invariance
\beq
w_4 1_2 + i \s_a w_{a} \to
e^{i\alpha} U (w_4 1_2 + i \s_a w_{a}) V, \quad
U \in SU(2)_1, \quad V \in SU(2)_2.
\label{221}
\eeq
The $T^{1,1}$ base is the intersection of this
with the $S^7 \in \Cbf^4$ of radius $r^{3/2}$:
\beq
\sum_i |w_i|^2 = r^3 \quad \Leftrightarrow 
\quad \tr (w_4 1_2 + i \s_a w_{a}) ( w_4 1_2 + i \s_a w_{a})^\dagger
= 2 r^3.
\eeq
This also has the invariance \myref{221},
demonstrating that $SU(2) \times SU(2) \times U(1)$
is an isometry of $T^{1,1}$.

On the other hand, when the embedding $z_1=\mu$
is imposed, we have a 4d real manifold $Y_4$ embedded
in the $\Cbf^3$ parameterized by $z_2,z_3,z_4$:
\beq
\mu z_2 = z_3 z_4.
\label{zem}
\eeq
This has a $U(1) \times U(1)$ invariance
with charges $(2,1,1)$ and $(0,1,-1)$ for the
three complex coordinates respectively.  There is
also a scaling symmetry $\Gamma$:
\beq
\Gamma : \quad z_2 \to \lambda^2 z_2, \quad
z_{3,4} \to \lambda z_{3,4}, \quad \lambda \in \Rbf_+.
\eeq
We declare the base of $Y_4$ to be $X_3 = Y_4/\Gamma$,
since any point in $Y_4$ can be reached from the
application of $\Gamma$ to a representative in $X_3$.
We can parameterize $Y_4$ by the pair $z_3,z_4$,
which it is useful to write as
\beq
z_3 = \rho e^{i \alpha} \cos \frac{\gamma}{2}, 
\quad z_4 = \rho e^{i \beta} \sin \frac{\gamma}{2},
\label{rs3}
\eeq
with $\gamma \in [0,\pi], \alpha,\beta \in [0,2\pi)$, and
$\rho \in [0,\infty)$.
This is just $\Cbf^2 = \Rbf_+ \times S^3$, or a family
of 3-spheres with radii $\rho$.
Note that \myref{zem} has a solution $z_2$
for every value of $z_3,z_4$, so that the entire
$\Rbf_+ \times S^3$ is contained in $Y_4$.  Also note
that for each value of $z_2$, there corresponds at least
one pair $z_3,z_4$.  Thus the entire $Y_4$
is parameterized by the $\Rbf_+ \times S^3$ \myref{rs3}.  
Alternatively, the base $X_3$ is the
intersection of $Y_4$ with any $S^3 \in \Cbf^2(z_3,z_4)$
corresponding to:
\beq
|z_3|^2 + |z_4|^2 = \rho^2.
\label{s3e}
\eeq
The equation \myref{zem} just tells us how the
$\Rbf_+ \times S^3$ parameterized by $\rho,\gamma,\alpha,\beta$
is embedded into $\Cbf^3(z_2,z_3,z_4)$.

Next note the homeomorphism determined by the continuous
deformation of~\myref{zem}:
\beq
\mu z_2 = (1-s) z_3 z_4, \quad s \in [0,1].
\eeq
At $s=1$ the embedding is just $z_2=0$ with
$z_3,z_4$ arbitrary.  Thus the topology of $Y_4$
is just $\Rbf_+ \times S^3$, and the topology
of $X_3$ is $S^3$.  The geometry of
$Y_4$ is different, since it is only the
projection into $\Cbf^2(z_3,z_4)$ that is
geometrically described by $\Rbf_+ \times S^3$, much as
an ellipse in 3d can be projected onto a
circle in a 2d plane.

It is of interest to relate $Y^4$
to the conifold geometry, particularly
the coordinate~$r$.  This relation follows from
\beq
r^3 = \sum_{i=1}^4 |z_i|^2 = \mu^2 + \rho^2 + \frac{1}{4\mu^2}
\rho^4 \sin^2 \gamma.
\eeq
First, note that as $r \to \mu^{2/3}$, the
$X_3 \simeq S^3$ radius $\rho$ shrinks to zero.  This is,
in detail, how the D7 brane ``ends'' in the $\adsv$
radial dimension.  Next note that if we fix
$r$, the $X_3 \simeq S^3$ radius becomes a function
of the polar angle $\gamma$.  The entire domain
of $\gamma$ has a solution, with $\rho(\gamma)$
falling in the range
\beq
2\mu (r^{3/2} - \mu) \leq \rho(\gamma) \leq r^3-\mu^2.
\eeq
The lower limit is saturated at $\gamma=\pi/2$,
whereas the upper limit is saturated at $\gamma=0,\pi$.
Thus at fixed $r$ the D7 embedding corresponds
to a 3d ellipsoid.  At $r \to \infty$ the ``squishing''
disappears and we just have an $S^3$.  This suggests
that a harmonic analysis at $r \to \infty$ in
terms of the coordinates $\gamma,\alpha,\beta$
should be relatively straightforward, involving
just the $S^3$ harmonics.

The $U(1) \times U(1)$ isometry of the 4d
manifold $Y_4$ is also an isometry of the 3d base,
as is apparent from \myref{s3e}.
This isometry group will be reflected in the spectrum
of eigenmodes and angular dependence.
A thorough harmonic analysis on this
3d space $X_3$ at arbitrary $\adsv$ radius $r$ is however, 
beyond the scope of the present article.

Finally consider the $\rhat \to \infty$ embedding
in terms of the conifold coordinates.
In this limit, the embedding is purely angular and is 
given by the sum:
\beq
X_3^\infty \equiv
\{ \theta_i,\phi_i,\psi ~ | ~ \theta_1=0, \psi=\phi_1+\phi_2 \}
+ \{ \theta_i,\phi_i,\psi ~ | ~ \theta_2=0, \psi=\phi_1+\phi_2 \}.
\label{rie}
\eeq
The two subspaces intersect at $\theta_1=\theta_2=0$.
Each subspace clearly contains an $S^2$ parameterized
by $(\theta_i,\phi_i)$, $i=1$ or $2$.  As a consequence,
an expansion on spherical harmonics $Y_{\ell m}(\theta_i,\phi_i)$
is valid, $i=1$ or $2$ depending on the subspace.
This affords a further justification for our 
assumption of $\tm=(\theta_1-\theta_2)/2$ 
independence in the $\rhat \to \infty$ limit for
the 8d scalars $\chi,\eta$.


\begin{thebibliography}{99}

\bibitem{Randall:1999ee}
L.~Randall and R.~Sundrum,
``A large mass hierarchy from a small extra dimension,''
Phys.\ Rev.\ Lett.\  {\bf 83} (1999) 3370
[arXiv:hep-ph/9905221].

\bibitem{Gherghetta:2006ha}
For a review, see T.~Gherghetta,
``Les Houches lectures on warped models and holography,''
arXiv:hep-ph/0601213.

\bibitem{Maldacena:1997re}
J.~M.~Maldacena,
``The large $N$ limit of superconformal field theories and supergravity,''
Adv.\ Theor.\ Math.\ Phys.\  {\bf 2} (1998) 231
[arXiv:hep-th/9711200].

\bibitem{Gubser:1998bc}
S.~S.~Gubser, I.~R.~Klebanov and A.~M.~Polyakov,
``Gauge theory correlators from non-critical string theory,''
Phys.\ Lett.\ B {\bf 428} (1998) 105
[arXiv:hep-th/9802109].

\bibitem{Witten:1998qj}
E.~Witten,
``Anti-de Sitter space and holography,''
Adv.\ Theor.\ Math.\ Phys.\  {\bf 2} (1998) 253
[arXiv:hep-th/9802150].

\bibitem{Klebanov:1998hh}
I.~R.~Klebanov and E.~Witten,
``Superconformal field theory on threebranes at a Calabi-Yau  singularity,''
Nucl.\ Phys.\ B {\bf 536} (1998) 199
[arXiv:hep-th/9807080].

\bibitem{Klebanov:2000hb}
I.~R.~Klebanov and M.~J.~Strassler,
``Supergravity and a confining gauge theory: Duality cascades and
$\chi$SB-resolution of naked singularities,''
JHEP {\bf 0008} (2000) 052
[arXiv:hep-th/0007191].

\bibitem{Giddings:2001yu}
S.~B.~Giddings, S.~Kachru and J.~Polchinski,
``Hierarchies from fluxes in string compactifications,''
Phys.\ Rev.\ D {\bf 66} (2002) 106006
[arXiv:hep-th/0105097].

\bibitem{Gubser:1998vd}
S.~S.~Gubser,
``Einstein manifolds and conformal field theories,''
Phys.\ Rev.\ D {\bf 59} (1999) 025006
[arXiv:hep-th/9807164].

\bibitem{Jatkar:1999zk}
D.~P.~Jatkar and S.~Randjbar-Daemi,
``Type IIB string theory on $AdS_5 \times T^{n,n'}$,''
Phys.\ Lett.\ B {\bf 460} (1999) 281
[arXiv:hep-th/9904187].

\bibitem{Ceresole:1999zs}
 A.~Ceresole, G.~Dall'Agata, R.~D'Auria and S.~Ferrara,
``Spectrum of Type IIB supergravity on $AdS_5 \times T^{1,1}$: 
predictions on $\Ncal = 1$ SCFT's,''
Phys.\ Rev.\ D {\bf 61} (2000) 066001
[arXiv:hep-th/9905226].

\bibitem{Ceresole:1999ht}
A.~Ceresole, G.~Dall'Agata and R.~D'Auria,
``KK spectroscopy of Type IIB supergravity on $AdS_5 \times T^{1,1}$,''
JHEP {\bf 9911} (1999) 009
[arXiv:hep-th/9907216].

\bibitem{Karch:2000gx}
A.~Karch and L.~Randall,
``Open and closed string interpretation of SUSY CFT's on branes with
boundaries,''
JHEP {\bf 0106} (2001) 063
[arXiv:hep-th/0105132].
  
\bibitem{Karch:2002sh}
A.~Karch and E.~Katz,
``Adding flavor to AdS/CFT,''
JHEP {\bf 0206} (2002) 043
[arXiv:hep-th/0205236].

\bibitem{Kruczenski:2003be}
M.~Kruczenski, D.~Mateos, R.~C.~Myers and D.~J.~Winters,
``Meson spectroscopy in AdS/CFT with flavour,''
JHEP {\bf 0307} (2003) 049
[arXiv:hep-th/0304032].

\bibitem{Levi:2005hh}
T.~S.~Levi and P.~Ouyang,
``Mesons and flavor on the conifold,''
arXiv:hep-th/0506021.

\bibitem{Breitenlohner:1982bm}
P.~Breitenlohner and D.~Z.~Freedman,
``Positive Energy In Anti-de Sitter Backgrounds And Gauged Extended
Supergravity,''
Phys.\ Lett.\ B {\bf 115} (1982) 197.

\bibitem{Breitenlohner:1982jf}
P.~Breitenlohner and D.~Z.~Freedman,
``Stability In Gauged Extended Supergravity,''
Annals Phys.\  {\bf 144} (1982) 249.
 
\bibitem{Arean:2004mm}
D.~Arean, D.~E.~Crooks and A.~V.~Ramallo,
``Supersymmetric probes on the conifold,''
JHEP {\bf 0411} (2004) 035
[arXiv:hep-th/0408210].

\bibitem{LOpri}
T.S. Levi and P. Ouyang, private communication.

\bibitem{Kim:1985ez}
H.~J.~Kim, L.~J.~Romans and P.~van Nieuwenhuizen,
``The Mass Spectrum Of Chiral $\Ncal=2$ $D=10$ Supergravity On $S^5$,''
Phys.\ Rev.\ D {\bf 32} (1985) 389.

\bibitem{Gunaydin:1984fk}
M.~Gunaydin and N.~Marcus,
``The Spectrum Of The $S^5$ Compactification Of The Chiral 
$\Ncal=2$, $D=10$
Supergravity And The Unitary Supermultiplets Of $U(2, 2/4)$,''
Class.\ Quant.\ Grav.\  {\bf 2} (1985) L11.

\bibitem{Brummer:2005sh}
F.~Brummer, A.~Hebecker and E.~Trincherini,
``The throat as a Randall-Sundrum model with Goldberger-Wise stabilization,''
Nucl.\ Phys.\ B {\bf 738} (2006) 283
[arXiv:hep-th/0510113].

\bibitem{Bergshoeff:1996tu}
E.~Bergshoeff and P.~K.~Townsend,
``Super D-branes,''
Nucl.\ Phys.\ B {\bf 490} (1997) 145
[arXiv:hep-th/9611173].

\bibitem{Grana:2002tu}
M.~Grana,
``D3-brane action in a supergravity background: The fermionic story,''
Phys.\ Rev.\ D {\bf 66} (2002) 045014
[arXiv:hep-th/0202118].

\bibitem{Marolf:2004jb}
D.~Marolf, L.~Martucci and P.~J.~Silva,
``The explicit form of the effective action for F1 and D-branes,''
Class.\ Quant.\ Grav.\  {\bf 21} (2004) S1385
[arXiv:hep-th/0404197].

\bibitem{Sakai:2003wu}
T.~Sakai and J.~Sonnenschein,
``Probing flavored mesons of confining gauge theories by supergravity,''
JHEP {\bf 0309} (2003) 047
[arXiv:hep-th/0305049].

\bibitem{Kuperstein:2004hy}
S.~Kuperstein,
``Meson spectroscopy from holomorphic probes on the warped deformed
conifold,''
JHEP {\bf 0503} (2005) 014
[arXiv:hep-th/0411097].

\end{thebibliography}
\end{document}